\documentclass[seceq,preprint]{ptptex}

\newcommand{\nn}{\nonumber\\}

\title{
Homotopy Operators and One-Loop Vacuum Energy at the Tachyon Vacuum%
}

\author{
Shoko \textsc{Inatomi},$^{1}$
Isao \textsc{Kishimoto}$^{2}$ and
Tomohiko \textsc{Takahashi}$^{1}$
}

\inst{
$^1$Department of Physics, Nara Women's University, Nara 630-8506,
Japan\\
$^2$Faculty of Education, Niigata University, Niigata 950-2181, Japan
}

\abst{
We construct homotopy operators for the BRST operator in the theory
around identity-based solutions, which are believed to represent the
tachyon vacuum in the cubic bosonic open string field theory.
Using the homotopy operators, we find that the one-loop vacuum energy at
the tachyon vacuum is independent of moduli such as interbrane
distances, which are included in the BRST operator.
We also revisit the cohomology problem, which was solved earlier without
the homotopy operators.
}

\begin{document}

\maketitle

\section{Introduction}

The cubic bosonic open string field theory (SFT) has classical solutions
describing 
the tachyon vacuum. 
In the theory expanded around the tachyon vacuum solution,
the BRST cohomology vanishes from a physical Hilbert space and
the annihilation of D-branes can be described using the solution.
However, there are apparently different results for the cohomology at
the tachyon vacuum.
For the solution constructed using wedge states,\cite{Schnabl:2005gv} 
the cohomology completely vanishes at all ghost
numbers.\cite{Ellwood:2006ba} \ 
On the other hand, it
was proved for the identity-based solution\cite{Takahashi:2002ez}
that there exists a
nonempty BRST cohomology at
unphysical ghost numbers.\cite{Kishimoto:2002xi} \ 
Also in numerical calculation,
while the cohomology was proposed to be
trivial,\cite{Ellwood:2001ig}
it was reported from a different analysis that the
cohomology exists at nonstandard ghost
numbers.\cite{Giusto:2003wc,Imbimbo:2006tz}

The vanishing cohomology for the wedge-like solution is proved using
a homotopy operator.\cite{Ellwood:2006ba} \ 
The anticommutator between the homotopy operator
and BRST operator at the solution
is equal to unity. Then, all the BRST closed states
turn out to be BRST exact.
In the case of the identity-based solution, the BRST operator 
at the solution
is represented as a well-defined operator acting on a
Fock space.\cite{Takahashi:2002ez} \ 
By a similarity transformation and a level shift operation in the ghost
sector, 
the cohomology can be obtained\cite{Kishimoto:2002xi}
from the known results for the ordinary
BRST operator.\cite{Kato:1982im,Henneaux:1986kp,Frenkel:1986dg}

In this work, we will construct homotopy operators for the
identity-based solutions.
We will find that the cohomologically nontrivial part is given by
acting the BRST operator to states,
which are obtained by acting the homotopy operator,
living {\it outside} a single Fock space.
Accordingly, although the resulting cohomology vanishes, there
is no contradiction with the earlier result since it was solved
only in a single Fock space.

Given the homotopy
operator, we cannot find nonzero scattering amplitudes with
on-shell external lines.
Then, it is natural to ask what happens to a one-loop vacuum amplitude
without external lines at the tachyon vacuum.
At the perturbative vacuum, the one-loop vacuum amplitude of an open string
depends on interbrane distances and it can be
interpreted as an amplitude for the exchange of a closed string between
D-branes. Since D-branes disappear at the tachyon vacuum, we can
speculate that the one-loop
vacuum amplitude becomes independent of interbrane distances.

To calculate the vacuum amplitude, gauge fixing is needed and it
seems a nontrivial problem at the tachyon vacuum.
However, the Siegel gauge works well for the theory around the
identity-based solution.
By using the Siegel gauge level expansion, the unstable
perturbative vacuum solution was found with high precision (up to the
truncation level (26,\,78)\cite{Kishimoto:2011zz})
and the one-loop vacuum amplitude was investigated
numerically.\cite{Takahashi:2011zz} \ 
Here, we will use the homotopy operator to analyze the vacuum amplitude
in the Siegel gauge and demonstrate the independence of interbrane
distances at the tachyon vacuum.

This paper is organized as follows. 
In \S \ref{sec:htp_op}, we briefly review 
the identity-based solution characterized by 
some function.
Then, we will explicitly construct the homotopy operator for 
a class of the identity-based solution.
In \S \ref{sec:1loop_ch},
we will evaluate the variation of the one-loop vacuum energy with
respect to moduli such as interbrane distances and we will discuss the
cohomology in the theory around the identity-based solution. We will
also comment on homotopy operators for other solutions.
Finally, we will give concluding remarks in \S \ref{sec:Rem}.

\section{Homotopy operators for the identity-based solution
\label{sec:htp_op}}

\subsection{Identity-based solution}

We consider the bosonic open string field theory with a midpoint
interaction. The action is given by
\begin{eqnarray}
 S[\Psi]&=&-\frac{1}{g^2}\int\left(\frac{1}{2}\Psi*Q_{\rm B}\Psi+
\frac{1}{3}\Psi*\Psi*\Psi\right),
\end{eqnarray}
where the operator $Q_{\rm B}$ is the Kato-Ogawa BRST charge, which is
defined on perturbative vacuum. The equation of motion is derived from
the variation of the action as
\begin{eqnarray}
\label{eq:eqom}
 Q_{\rm B}\Psi+\Psi*\Psi=0.
\end{eqnarray}

We can construct an exact classical solution of the equation of
motion~(\ref{eq:eqom}) using half-string operators and the identity
string field $I$:\cite{Takahashi:2002ez}
\begin{eqnarray}
\label{eq:sol}
 \Psi_0&=&Q_L(e^h-1)I-C_L((\partial h)^2 e^h)I,
\end{eqnarray}
where $Q_L(f)$ and $C_L(f)$ are integrations of the BRST current
$j_{\rm B}(z)$ and the ghost $c(z)$ with a function $f(z)$ along a
half-unit disc:
\begin{eqnarray}
 Q_L(f)=\int_{C_{\rm left}}\frac{dz}{2\pi i}f(z)j_{\rm B}(z),\ \ \ 
 C_L(f)=\int_{C_{\rm left}}\frac{dz}{2\pi i}f(z)c(z).
\end{eqnarray}
We can find that the equation of motion holds for the function $h(z)$
such that $h(-1/z)=h(z)$ and $h(\pm i)=0$. Expanding the string field as
$\Psi=\Psi_0+\Phi$
and subtracting $S[\Psi_0]$, 
we obtain the action for the fluctuation $\Phi$
around the solution $\Psi_0$ as
\begin{eqnarray}
\label{eq:actionPhi}
  S'[\Phi]&=&-\frac{1}{g^2}\int\left(\frac{1}{2}\Phi*Q'\Phi+
\frac{1}{3}\Phi*\Phi*\Phi\right).
\end{eqnarray}
The operator $Q'$ in the quadratic term is given by
\begin{eqnarray}
\label{eq:Q}
 Q'&=& Q(e^h)-C((\partial h)^2 e^h),
\end{eqnarray}
where the operators $Q(f)$ and $C(f)$ are defined as
integrations along a unit circle:
\begin{eqnarray}
 Q(f)=\oint \frac{dz}{2\pi i}f(z)j_{\rm B}(z),\ \ \ 
 C(f)=\oint \frac{dz}{2\pi i}f(z)c(z).
\end{eqnarray}

The classical solution (\ref{eq:sol}) includes an arbitrary function,
which is changed by gauge transformations. 
Most of the solutions are regarded as a trivial pure gauge
transformation from the trivial solution, $\Psi_0=0$, but nontrivial
solutions can be obtained at the boundary of some function spaces.
For example, we consider the classical solution constructed using the
function
\begin{eqnarray}
\label{eq:haz}
 h_a(z)&=& \log\left(1+\frac{a}{2}\left(z+z^{-1}\right)^2\right).
\end{eqnarray}
This function includes one parameter $a$, which is larger than or equal
to $-1/2$. This range of the parameter $a$ is determined with the reality
condition for the classical solution.
The solution for $h_a(z)$ corresponds to a trivial pure
gauge for $a>-1/2$, but we find that it becomes nontrivial at
the boundary $a=-1/2$. 
In the case of $a=-1/2$, the operator ($\ref{eq:Q}$) has no cohomology
in the ghost number one sector.\cite{Kishimoto:2002xi} \ 
In addition, the theory based on the action ($\ref{eq:actionPhi}$) has
an unstable vacuum solution corresponding to the perturbative string
vacuum.\cite{Kishimoto:2011zz,Kishimoto:2009hc,Kishimoto:2009nd} \ 
Consequently, we conclude that the nontrivial
solution at $a=-1/2$ is the tachyon vacuum solution.

The function (\ref{eq:haz}) for the solution (\ref{eq:sol})
can be generalized as
\begin{eqnarray}
\label{eq:halz}
 h_a^l(z)&=&\log\left(1-\frac{a}{2}(-1)^l(z^l-(-1)^lz^{-l})^2\right),
\end{eqnarray}
where $l$ is a positive integer
and $a\geq -1/2$.\cite{Kishimoto:2002xi} \ 
This includes the function (\ref{eq:haz}) as the $l=1$ case.
For all $l$, the classical solution is expected to be the tachyon vacuum
solution at $a=-1/2$.
For the function $h_{-1/2}^l(z)$, 
the BRST operator (\ref{eq:Q}) at the tachyon vacuum
can be written as
\begin{eqnarray}
&&Q_l=Q(F)+C(G),
\label{eq:Qldef}
\\
&&F(z)=\frac{(-1)^l}{4}\left(z^l+(-1)^lz^{-l}\right)^2,~~ 
 G(z)=-{(-1)^ll^2}z^{-2}\left(z^l-(-1)^lz^{-l}\right)^2. 
\label{eq:FGdef}
\end{eqnarray}

\subsection{Homotopy operators
\label{sec:Hop}}

The operator product expansion (OPE) of the BRST current with the
antighost $b(z)$ is
\begin{eqnarray}
\label{eq:opejb}
 j_{\rm B}(z)b(z')\sim \frac{3}{(z-z')^3}+\frac{1}{(z-z')^2}j_{\rm
  gh}(z') +\frac{1}{z-z'}T(z'),
\end{eqnarray}
where $j_{\rm gh}(z)$ is the ghost number current and $T(z)$ is the
total energy momentum tensor. From this OPE, we can derive the
anticommutation relation of $Q(f)$ and $b(z)$ as
\begin{eqnarray}
\label{eq:Qfb}
 \{Q(f),\,b(z)\}&=&
\frac{3}{2}\partial^2 f(z)+\partial f(z)\,j_{\rm gh}(z)
+f(z)\,T(z).
\end{eqnarray}
Similarly, from the OPE of $c(z)$ and $b(z)$, we obtain the
anticommutation relation,
\begin{eqnarray}
\label{eq:Cfb}
 \{C(f),\,b(z)\}&=& f(z).
\end{eqnarray}

Using (\ref{eq:Qfb}) and (\ref{eq:Cfb}),
we can calculate the anticommutation relation of the BRST operator
(\ref{eq:Qldef}) with $b(z)$.
The important point is that the function $F(z)$ in (\ref{eq:FGdef})
has second-order zeros at the points,
\begin{eqnarray}
 z_k&=&\left\{
\begin{array}{ll}
\displaystyle
 e^{i\frac{k-1}{l}\pi} & {\rm for}\ {\rm odd}\ l,\\
\displaystyle
 e^{i\frac{2k-1}{2l}\pi} & {\rm for}\ {\rm even}\ l,
\end{array}
\right.\ \ \ 
(k=1,2,\cdots,2l)
\end{eqnarray}
which are solutions to the equation: $z^{2l}+(-1)^l=0$.
Therefore, the anticommutator becomes a c-number:
\begin{eqnarray}
\label{eq:Qlb}
 \{Q_l,\,b(z_k)\}&=&
\frac{3}{2}\partial^2 F(z_k)+G(z_k)=
z_k^{-2}l^2.
\end{eqnarray}
It should be noted that the function $e^{h^l_a}$ has only first-order
zeros for $a> -1/2$, and therefore, the above anticommutator depends on
the ghost number current $j_{\rm gh}$ for trivial pure gauge solutions.

Alternatively, we can explicitly compute
the anticommutation relation (\ref{eq:Qlb}) in
terms of the oscillator expression. The operator $Q_l$ is expanded as
\begin{eqnarray}
\label{eq:Qlexp}
 Q_l&=& \frac{1}{2}Q_{\rm B}+\frac{(-1)^l}{4}(Q_{2l}+Q_{-2l})
+2l^2 c_0 -(-1)^l l^2 (c_{2l}+c_{-2l}),
\end{eqnarray}
where we have expanded the BRST current and the ghost as
$j_{\rm B}(z)=\sum_n Q_n z^{-n-1}$ and $c(z)=\sum_n c_n z^{-n+1}$.
Using the oscillator expressions of $b(z)=\sum_n b_n
z^{-n-2}$, $T(z)=\sum_nL_n z^{-n-2}$ and $j_{\rm gh}(z)=\sum_n q_n
z^{-n-1}$, we find the anticommutation relation of $Q_m$ and $b_n$ from
(\ref{eq:opejb}),
\begin{eqnarray}
\label{eq:Qmbn}
 \{Q_m,\,b_n\}&=& L_{m+n}+m q_{m+n}+\frac{3}{2}m(m-1)\delta_{m+n,0}.
\end{eqnarray}
Using (\ref{eq:Qlexp}) and (\ref{eq:Qmbn}), we can calculate the
left-hand side of (\ref{eq:Qlb}) as
\begin{eqnarray}
 \{Q_l,\,b(z_k)\}&=&\frac{1}{2}\sum_{n=-\infty}^\infty L_n(z_k)^{-n-2}
\nn
&&
+\frac{(-1)^l}{4}\left(
\sum_{n=-\infty}^\infty(L_{n+2l}+2lq_{n+2l})(z_k)^{-n-2}
+\frac{3}{2}2l(2l+1)(z_k)^{-2l-2}
\right)
\nn
&&
+\frac{(-1)^l}{4}\left(
\sum_{n=-\infty}^\infty(L_{n-2l}-2lq_{n-2l})(z_k)^{-n-2}
+\frac{3}{2}2l(2l-1)(z_k)^{2l-2}
\right)
\nn
&&
+2l^2 (z_k)^{-2}-(-1)^l l^2 ((z_k)^{-2l-2}+(z_k)^{2l-2})
\nn
&=& z_k^{-2}l^2,
\end{eqnarray}
where we have used $(z_k)^{2l}=-(-1)^l$.
Thus, Eq.~(\ref{eq:Qlb}) can be derived from
the mode expansion without any divergence.

The anticommutation relation (\ref{eq:Qlb}) implies that we can define
a homotopy operator $\hat A$ corresponding to the BRST operator
$Q_l$ at the solution (\ref{eq:sol}) with the function (\ref{eq:halz})
at $a=-1/2$:
\begin{eqnarray}
&&\hat A= \sum_{k=1}^{2l}\,a_k\,l^{-2}z_k^2\,b(z_k),
~~~~~~~\sum_{k=1}^{2l}\,a_k =1,
\label{eq:hatAdef}
\end{eqnarray}
which satisfies the relations
\begin{eqnarray}
&&\{Q_l,\hat A\}=1,~~~~~\hat A^2=0.
\label{eq:htp_rel}
\end{eqnarray}
If we choose the coefficients $a_k$ as
\begin{eqnarray}
 a_k=a_{l-k+2},~~(k=1,2,\cdots,l+1);~~~~~
a_k=a_{3l-k+2},~~(k=l+2,l+3,\cdots,2l)
\label{eq:BPZlodd}
\end{eqnarray}
for odd $l$,
\begin{eqnarray}
 a_k=a_{l-k+1},~~(k=1,2,\cdots,l);~~~~~
a_k=a_{3l-k+1},~~(k=l+1,l+2,\cdots,2l)
\label{eq:BPZleven}
\end{eqnarray}
for even $l$ and $a_k\in \mathbb{R}\,(k=1,2,\cdots, 2l)$, the
operator $\hat A$ is BPZ even and Hermitian.
(The conditions (\ref{eq:BPZlodd}) and (\ref{eq:BPZleven}) 
imply that the coefficients in (\ref{eq:hatAdef}) 
corresponding to each pair of $z_k$, which are symmetric points
with respect to the imaginary axis, are equal.)
In this case, $\hat A$ is 
explicitly expressed in terms of oscillators as
\begin{eqnarray}
\hat A=l^{-2}\sum_{m=-\infty}^{\infty}\sum_{k=1}^{2l}
a_k\left(b_{2m}\cos\frac{2m(k-1)\pi}{l}
-ib_{2m-1}\sin\frac{(2m-1)(k-1)\pi}{l}
\right)
\end{eqnarray}
for odd $l$ and
\begin{eqnarray}
\hat A=l^{-2}\sum_{m=-\infty}^{\infty}\sum_{k=1}^{2l}
a_k\left(b_{2m}\cos\frac{2m(2k-1)\pi}{2l}
-ib_{2m-1}\sin\frac{(2m-1)(2k-1)\pi}{2l}
\right)
\end{eqnarray}
for even $l$.

Note that the above homotopy operator $\hat A$ can be rewritten as
\begin{eqnarray}
 \hat{A}\Phi=\frac{1}{2}\left(
A*\Phi+(-1)^{|\Phi|}\Phi*A
\right),
\label{eq:op_state}
\end{eqnarray}
using the homotopy state $A\equiv \hat A I$.
In order to obtain this expression, we have used (\ref{eq:BPZlodd}),
(\ref{eq:BPZleven}) and 
\begin{eqnarray}
\label{eq:bstar}
 z^{-4}b(-1/z)\Phi_1*\Phi_2&=&
   (-1)^{|\Phi_1|}\Phi_1*b(z)\Phi_2,\\
\label{eq:bid}
 z^{-4}b(-1/z)I&=&
   b(z)I.
\end{eqnarray}
Using  $Q_l I=0$ and (\ref{eq:htp_rel}), we have
\begin{eqnarray}
&&Q_lA=I,~~~~~~~~~A*A=0
\label{eq:QA=I}
\end{eqnarray}
for the homotopy state $A$.

\section{One-loop vacuum energy and cohomology
\label{sec:1loop_ch}}

\subsection{One-loop vacuum energy at the tachyon vacuum
\label{sec:1loop}}

We consider a string field theory at the tachyon vacuum, in which the
BRST operator is given by $Q_l$. We impose the Siegel gauge condition,
$b_0\Phi=0$. Then, the one-loop vacuum energy is given by the
integration over the moduli $t$ of the partition function:
\begin{eqnarray}
\label{eq:partfunc}
 Z(t)&=& {\rm Tr}\left[(-1)^{N_{\rm FP}}e^{-t L'}b_0c_0\right].
\end{eqnarray}
Here, $L'=\{Q_l,\,b_0\}$ is the Siegel gauge inverse
propagator\footnote{There are several works about the Siegel gauge
theory around the identity-based solution. In
Ref.~\citen{Drukker:2003hh}, it was suggested  that
purely closed string amplitudes could be derived from open string
fields by using the kinetic operator $L'$. It was found in
Ref.~\citen{Takahashi:2003xe} that all the scattering amplitudes
vanish. The vacuum structure in the theory with $L'$
was numerically evaluated up to level
26\cite{Kishimoto:2009hc,Kishimoto:2009nd}, and the resulting structure
agrees with that of the tachyon vacuum.}
and
$N_{\rm FP}$ is the operator counting ghost number:
$N_{\rm FP}=c_0b_0+\sum_{n\geq
1}(c_{-n}b_n-b_{-n}c_n)$.
The trace {\rm Tr} is defined by the sum over all the Fock space states
and the projection operator $b_0c_0$ is inserted into the trace to
restrict to the Siegel gauge subspace.

Suppose that there are multiple separated D-branes at the perturbative
vacuum. 
Then, the BRST operator $Q_{\rm B}$  depends on interbrane distances
through the zero mode of string coordinates in the nondiagonal sector
of a string field with the Chan-Paton indices.\footnote{To introduce
Wilson lines in the theory on D25 branes, 
we have only to replace
momentum zero modes of string coordinates as $p^m\rightarrow p^m+
(\theta_i-\theta_j)/\pi R$, where $i$ and $j$ are Chan-Paton
indices.\cite{polchinskiI} \  Therefore, in the T-dual picture, 
we can describe separated multiple D-branes only by changing zero modes
of string coordinates in the SFT action of coincident multiple
D-branes\cite{polchinskiI}.}
In this case, $\Psi_0$ (\ref{eq:sol})  with the function
$h=h^l_{a=-1/2}$ is a solution by including the identity matrix 
in the identity state $I$
 and it is expected to represent the tachyon vacuum.
The partition function (\ref{eq:partfunc}) includes 
the trace over the Chan-Paton indices and apparently depends on the
brane distance parameters through $L'$ or
the BRST operator at the solution $Q_l$.
However, the one-loop vacuum energy is expected not to
include the distance parameters, since the D-branes no longer
exist at the tachyon vacuum.

Let us show that (\ref{eq:partfunc}) is indeed independent of the
interbrane distances. Under an infinitesimal change of modulus such as
the D-brane
positions, the BRST operator changes to $Q_l'=Q_l+\delta Q_l$.
The variation of $L'$ is given by that of the BRST
charge:
\begin{eqnarray}
 \delta L'=\{\delta Q_l,\,b_0\}.
\end{eqnarray}
The key ingredient of the proof is the existence of the homotopy
operator $\hat{A}$. Since the homotopy operator (\ref{eq:hatAdef}) is
defined only by the antighost, $\hat{A}$ and $b_0$ 
anticommute with each other:
\begin{eqnarray}
\label{eq:Ab0}
 \{\hat{A},\,b_0\}=0.
\end{eqnarray}
From the variation of (\ref{eq:htp_rel}), we find
\begin{eqnarray}
\label{eq:QlA}
 \{\delta Q_l,\,\hat{A}\}=0.
\end{eqnarray}
Using the Jacobi identity, we have
\begin{eqnarray}
\label{eq:LA}
[L',\,\hat{A}]=[\{Q_l,\,b_0\},\,\hat{A}]=
-[\{b_0,\,\hat{A}\},\,Q_l]-[\{\hat{A},\,Q_l\},\,b_0]=0.
\end{eqnarray}

Now, we are ready to evaluate the change of the partition function:
\begin{eqnarray}
\label{eq:dZ1}
 \delta Z(t)=-t \int_0^1 d\alpha\,{\rm Tr}\left[(-1)^{N_{\rm FP}}
e^{-\alpha t L'} \{\delta Q_l,\,b_0\} e^{-(1-\alpha) t L'}b_0 c_0
\right].
\end{eqnarray}
Using the commutation relations $[L',\,b_0]=0$
and the cyclic invariance of the
trace, we can rewrite the integrand in (\ref{eq:dZ1}) as
\begin{eqnarray}
&&
 {\rm Tr}\left[(-1)^{N_{\rm FP}}
e^{-\alpha t L'} b_0\,\delta Q_l\,e^{-(1-\alpha) t L'}b_0 c_0
\right]\nn
&=&
- {\rm Tr}\left[(-1)^{N_{\rm FP}}
e^{-\alpha t L'} \,\delta Q_l\,e^{-(1-\alpha) t L'}b_0 c_0 b_0
\right]\nn
&=&
- {\rm Tr}\left[(-1)^{N_{\rm FP}}
e^{-\alpha t L'} \,\delta Q_l\,e^{-(1-\alpha) t L'}b_0
\right].
\end{eqnarray}
In this equation, we insert $\{Q_l,\,\hat{A}\}(=1)$ between $e^{-(1-\alpha)
t L'}$ and $b_0$:
\begin{eqnarray}
 &=&
- {\rm Tr}\left[(-1)^{N_{\rm FP}}
e^{-\alpha t L'} \,\delta Q_l\,e^{-(1-\alpha) t L'}\{Q_l,\,\hat{A}\}b_0
\right]\nn
 &=&
- {\rm Tr}\left[(-1)^{N_{\rm FP}}
e^{-\alpha t L'} \,\delta Q_l\,e^{-(1-\alpha) t L'}Q_l\,\hat{A}\,b_0
\right]\nn
&&
\label{eq:dZ2}
\ \ \ \ \ - {\rm Tr}\left[(-1)^{N_{\rm FP}}
e^{-\alpha t L'} \,\delta Q_l\,e^{-(1-\alpha) t L'}\hat{A}\,Q_l\,b_0
\right].
\end{eqnarray}
In the second term, we move $\hat A$ to the left using
the (anti) commutation relations,
 (\ref{eq:QlA}) and (\ref{eq:LA}).
Then, we have
\begin{eqnarray}
  &=&
- {\rm Tr}\left[(-1)^{N_{\rm FP}}
e^{-\alpha t L'} \,\delta Q_l\,e^{-(1-\alpha) t L'}Q_l\,\hat{A}\,b_0
\right]\nn
&&
\ \ \ \ \ - {\rm Tr}\left[(-1)^{N_{\rm FP}}
e^{-\alpha t L'} \,\delta Q_l\,e^{-(1-\alpha) t L'}\,Q_l\,b_0\hat A
\right],
\end{eqnarray}
using cyclic invariance of the trace.
These two terms cancel each other, thanks to
(\ref{eq:Ab0}).
Thus, we finally obtain
\begin{eqnarray}
 \delta Z(t)=0,
\label{eq:dZt=0}
\end{eqnarray}
and we conclude that the one-loop vacuum
energy is independent of interbrane distances.

We note that the above proof of $\delta Z(t)=0$ does not depend on the
details of the variation of $Q_l$ and the calculation is rather formal.
If we restrict the space in the definition of the trace, 
it is necessary to treat the relation $\{Q_l,\hat A\}=1$
carefully. We will discuss the related issue in the next subsection.

\subsection{Homotopy operator versus cohomology
\label{sec:vs}}

Once the homotopy operator $\hat A$ exists at the tachyon vacuum,
a BRST invariant state $\psi$ such that $Q_l\psi=0$ 
is a BRST exact state, namely,
\begin{eqnarray}
\label{eq:cohomology1}
 Q_l\psi=0\ \ \ &\Leftrightarrow&\ \ \  \psi=Q_l(\hat A\psi),
\end{eqnarray}
because of the commutation relation $\{Q_l,\hat A\}=1$
(\ref{eq:htp_rel}).

On the other hand, the cohomology of the BRST operator $Q_l$ was derived
earlier in Ref.~\citen{Kishimoto:2002xi} by referring to the results for
$Q_{\rm B}$\cite{Kato:1982im,Henneaux:1986kp,Frenkel:1986dg}:
\begin{eqnarray}
\label{eq:cohomology2}
 Q_l\psi=0\ \ \ &\Leftrightarrow&
\ \ \  |\psi\rangle=|P\rangle\otimes U_l\,b_{-2l}b_{-2l+1}\cdots
b_{-2}|0\rangle\nn
&&\hspace{1.5cm}
+|P'\rangle\otimes U_l\,b_{-2l+1}b_{-2l+2}\cdots b_{-2}|0\rangle
+Q_l|\phi\rangle,
\end{eqnarray}
where $|P\rangle$ and $|P'\rangle$ are positive-norm states in the
matter sector such as DDF states, and the operator $U_l$ is given by
\begin{eqnarray}
\label{eq:Ul}
 U_l=\exp\left(-2\sum_{n=1}^\infty \frac{(-1)^{n(l+1)}}{n}q_{-2nl}\right).
\end{eqnarray}
According to the result of (\ref{eq:cohomology2}), the cohomology exists
in the Hilbert space of the ghost numbers $-2l+1$ and $-2l+2$. This
result is apparently
incompatible with vanishing cohomology 
in all the ghost number sectors,
which can be read off from (\ref{eq:cohomology1}). \\

In order to resolve the discrepancy between (\ref{eq:cohomology1})
and (\ref{eq:cohomology2}), 
we investigate the cohomologically
nontrivial states in (\ref{eq:cohomology2})
\begin{eqnarray}
 |\varphi\rangle &=&|P\rangle\otimes U_l\,b_{-2l}b_{-2l+1}\cdots
b_{-2}|0\rangle
+|P'\rangle\otimes U_l\,b_{-2l+1}b_{-2l+2}\cdots b_{-2}|0\rangle.
\label{eq:varphidef}
\end{eqnarray}
According to the proposition (\ref{eq:cohomology1}), we can represent 
the state $|\varphi\rangle$ as a BRST exact state:
\begin{eqnarray}
\label{eq:QAvarphi}
 |\varphi\rangle =Q_l(\hat{A}|\varphi \rangle).
\end{eqnarray}
Here, let us rewrite the state $\hat{A}|\varphi\rangle$ using a ``normal
ordered'' expression, namely moving $\hat{A}$, which includes the
positive frequency modes, to the right of $U_l$.
Using the commutation relation $[q_m,\,b_n]=-b_{m+n}$, we have
\begin{eqnarray}
 b(z)\,U_l &=& \exp\left(-2\sum_{n=1}^\infty \frac{(-1)^{n(l+1)}}{n}
z^{-2nl}\right)\,U_l\,b(z),
\end{eqnarray}
for $U_l$ given in (\ref{eq:Ul}).
By taking a limit, $z\to z_k$ with $(z_k)^{2l}=-(-1)^l$,
\begin{eqnarray}
 b(z_k)\,U_l b_{-m}\cdots b_{-2}|0\rangle
 = \exp\left(-2\sum_{n=1}^\infty \frac{1}{n}
\right)\,U_l\,b(z_k) b_{-m}\cdots b_{-2}|0\rangle
 =0
\label{eq:bzkUl=0}
\end{eqnarray}
is obtained.
Because the homotopy operator $\hat A$  (\ref{eq:hatAdef})
is a linear combination of $b(z_k)$,
this implies that the state $\hat{A}|\varphi\rangle$ 
becomes zero in the Fock space expression.
Namely, all the coefficients of $\hat{A}|\varphi\rangle$
vanish when it is expanded in terms of the conventional oscillators,
$b_{-n},c_{-m}$, on the conformal vacuum $|0\rangle$.
However, the relation (\ref{eq:QAvarphi}) implies that the state
$\hat{A}|\varphi\rangle$ is not truly zero in some enlarged space.
The conclusion is that the state $|\varphi\rangle$ is an exact state
obtained by applying $Q_l$ to the state outside a single Fock space,\footnote{
In this paper, we regard a space in which any state can be expressed by a
linear combination of Fock bases with finite coefficients
as a ``single Fock space'', where  a single Fock vacuum
and a single set of creation-annihilation operators are fixed.
(In the present case, they correspond to $c_1|0\rangle$
and $\{b_n,c_m\}$ in the ghost sector, respectively.
Conversely, we bear in mind the Bogoliubov transformation, for example,
as outside of a single Fock space.)
Although we should define the space as a completion with respect to 
appropriate norm mathematically, we leave it as a future problem.
Intuitively, we may be able to interpret that
$\hat A|\varphi\rangle$ weakly converges to zero but not in
a strong sense.
We can find a similar situation with respect to the ``phantom term''
$\psi_{\infty}$ in Ref.~\citen{Schnabl:2005gv}.
} which is beyond the scope of the proposition (\ref{eq:cohomology2}).

\subsection{Comments on homotopy operators 
for other solutions}

Firstly, we comment on another type of
identity-based solution,\cite{Igarashi:2005wh}
which is given by a function $h^{(4)}_a$
with a parameter $a\geq -1/2$:
\begin{eqnarray}
 h^{(4)}_a(z)&=&\log\left(1+2a-\frac{a}{8}\left(z-z^{-1}\right)^4
\right)
\end{eqnarray}
for $h$ in (\ref{eq:sol}).
At the boundary of the parameter, i.e., $a=-1/2$, the solution is expected to
represent a nontrivial solution.
For the function $h^{(4)}_{-1/2}$, the BRST operator around the solution
is
\begin{eqnarray}
&&Q^{(4)}=Q(F_4)+C(G_4)\nn
&&~~~~~=\frac{3}{8}Q_{\rm B}-\frac{1}{4}(Q_2+Q_{-2})
+\frac{1}{16}(Q_4+Q_{-4})+2c_0-c_4-c_{-4},
\label{eq:Q4def}
\\
&&F_4(z)=\frac{1}{16}\left(z-z^{-1}\right)^4,~~~
G_4(z)=-z^{-2}(z^2-z^{-2})^2.
\end{eqnarray}
In this case, the function $F_4(z)$ has fourth-order zeros at 
$z=\pm 1$. Using the anticommutation relations,
 (\ref{eq:Qfb}) and (\ref{eq:Cfb}), we have
\begin{eqnarray}
&&\{Q^{(4)},b(\pm 1)\}=0,~~~~~\{Q^{(4)},\partial b(\pm 1)\}=0,\\
&&\{Q^{(4)},\partial^2 b(\pm 1)\}=\frac{3}{2}\partial^4 F_4(\pm 1)
+\partial^2 G_4(\pm 1)=4.
\end{eqnarray}
Therefore, we obtain a homotopy operator $\hat A^{(4)}$:
\begin{eqnarray}
\hat A^{(4)}&=&\frac{1}{8}(\partial^2 b(1)+\partial^2 b(-1))
+\frac{5}{8}(\partial b(1)-\partial b(-1))
+\frac{1}{2}(b(1)+b(-1))\nn
&=&\sum_{n=-\infty}^{\infty}n^2\,b_{2n}
\label{eq:hatA4}
\end{eqnarray}
which is BPZ even and Hermitian and satisfies
\begin{eqnarray}
 \{Q^{(4)},\hat A^{(4)}\}=1,~~~~(\hat A^{(4)})^2=0.
\label{eq:htp_rel4}
\end{eqnarray}
The first anticommutation relation can also be obtained using mode
expansion from  (\ref{eq:Q4def}), (\ref{eq:hatA4}), and (\ref{eq:Qmbn}).
Noting the relation $\{\hat A^{(4)},b_0\}=0$, we can trace the same 
computation as \S \ref{sec:1loop} to demonstrate 
$\delta Z(t)=0$ (\ref{eq:dZt=0}).

In general, we can construct classical solutions 
with higher-order zeros.\cite{Igarashi:2005wh} \ 
If the function corresponding to the solution has an $n$-th order zero
at $z=z_0$,
the anticommutator between $\partial^{k-2} b(z)$ and the BRST operator
at the solution becomes a nonzero c-number at $z=z_0$ for $k=n$ and
it vanishes for $k<n$.
Using these anti-commutation relations, the homotopy operator can be
obtained as a linear combination of $\partial^{k-2} b(z)\ (k=2,\cdots,n)$.
Therefore, we expect that the classical solution with higher-order zeros
corresponds to the tachyon vacuum.

The cohomology of $Q^{(4)}$ was derived in Ref.~\citen{Igarashi:2005wh}
and the result is
\begin{eqnarray}
\label{eq:cohomology2_4}
 Q^{(4)}\psi=0\ \ \ &\Leftrightarrow&
\ \ \  |\psi\rangle=|P\rangle\otimes U_{(4)}\,b_{-4}b_{-3}b_{-2}|0\rangle\nn
&&\hspace{1.5cm}
+|P'\rangle\otimes U_{(4)}\,b_{-3}b_{-2}|0\rangle
+Q^{(4)}|\phi\rangle,
\end{eqnarray}
in a similar way to (\ref{eq:cohomology2}), where
\begin{eqnarray}
 U_{(4)}&=&\exp\left(-4\sum_{n=1}^{\infty}\frac{1}{n}q_{-2n}\right).
\end{eqnarray}
Namely, in the Fock space expression, a state of the form
\begin{eqnarray}
{}|\varphi_{(4)}\rangle&=&|P\rangle\otimes
 U_{(4)}\,b_{-4}b_{-3}b_{-2}|0\rangle
+|P'\rangle\otimes U_{(4)}\,b_{-3}b_{-2}|0\rangle
\end{eqnarray}
represents a nontrivial state of $Q^{(4)}$ cohomology.
However, from (\ref{eq:htp_rel4}), it could be rewritten as
\begin{eqnarray}
 {}|\varphi_{(4)}\rangle&=&Q^{(4)}({\hat A}^{(4)}|\varphi_{(4)}\rangle).
\label{eq:varphi_exact4}
\end{eqnarray}
This apparent inconsistency can be resolved as in \S \ref{sec:vs}.
Noting the relations
\begin{eqnarray}
b(z) U_{(4)}&=&
\exp\left(-4\sum_{n=1}^{\infty}\frac{z^{-2n}}{n}\right)U_{(4)}b(z) 
=(1-z^{-2})^4\,U_{(4)}b(z),\\
\partial b(z) U_{(4)}&=&(1-z^{-2})^4\,U_{(4)}\partial b(z)+8z^{-3}
(1-z^{-2})^3\,U_{(4)}b(z),\\
\partial^2 b(z) U_{(4)}&=&(1-z^{-2})^4\,U_{(4)}\partial^2 b(z)+16z^{-3}
(1-z^{-2})^3\,U_{(4)}\partial b(z)
\nn
&&-24z^{-4}(1-3z^{-2})(1-z^{-2})^2\,U_{(4)}b(z),
\end{eqnarray}
and taking a limit $z\to \pm 1$, we have
\begin{eqnarray}
&&b(\pm 1)\,U_{(4)} b_{-m}\cdots b_{-2}|0\rangle=0,~~~~~
\partial b(\pm 1)\,U_{(4)} b_{-m}\cdots b_{-2}|0\rangle=0,
\nn
&&\partial^2 b(\pm 1)\,U_{(4)} b_{-m}\cdots b_{-2}|0\rangle=0,
\end{eqnarray}
by reexpressing the left-hand sides as normal ordered forms.
Because the homotopy operator ${\hat A}_{(4)}$ 
is given as a linear combination of
$b(\pm 1),\partial b(\pm 1)$ and $\partial^2b(\pm 1)$
as in (\ref{eq:hatA4}), the above equations imply that
$\hat A_{(4)}|\varphi_{(4)}\rangle$ 
is zero in the Fock space expression.
Hence, it is necessary to use the appropriate expression beyond 
a single Fock space to conclude that
$|\varphi_{(4)}\rangle$ is $Q^{(4)}$ exact
in the sense of (\ref{eq:varphi_exact4}).\\

Next, we briefly mention the case of solutions
constructed using the $KBc$ subalgebra.
The Schnabl solution\cite{Schnabl:2005gv} and Erler-Schnabl
solution\cite{Erler:2009uj} are in this category
 and they are considered to represent the tachyon vacuum.
The homotopy states, which satisfy (\ref{eq:QA=I}),
for the BRST operator around these solutions
were obtained in Refs.~\citen{Ellwood:2006ba} 
and~\citen{Erler:2009uj}.
These states can be rewritten as the homotopy operators such as
(\ref{eq:htp_rel}) through the definition given by
(\ref{eq:op_state}).\cite{Ellwood:2006ba} \ 
It turns out that both of them do not anticommute with $b_0$.
Hence, we cannot apply the same procedure as in
\S\ref{sec:1loop} for these solutions.\footnote{In the theory around the
{\it regularized} identity-based
solution,\cite{Arroyo:2010fq, Zeze:2010sr,Arroyo:2010sy} we cannot
apply the procedure for the same reason.
Although we have used the Siegel gauge condition for the evaluation of
the one-loop vacuum energy, other gauge
conditions\cite{Kiermaier:2008jy, Asano:2008iu} might be useful for
these solutions.
} 

However, for a solution $\Psi=\sqrt{1-\beta K}\beta^{-1}c\sqrt{1-\beta
K}$, which is a real form of an identity-based solution: $\beta^{-1} c-cK$,
in terms of the $KBc$ subalgebra, the homotopy state is $A=\beta B$ as
is given in Ref.~\citen{Schnabl:2010tb}.
Therefore, the homotopy operator $\hat A$ such as (\ref{eq:htp_rel}),
which is a linear combination of $b_n$,
satisfies (\ref{eq:Ab0}) and, therefore, the same computation 
in \S\ref{sec:1loop} can be applicable to prove $\delta Z(t)=0$
(\ref{eq:dZt=0}).

\section{Concluding remarks
\label{sec:Rem}}

In this work, we have constructed a homotopy operator $\hat A$
for the BRST operator $Q_l$ in the theory around 
a type of identity-based solution associated with  
particular functions $h^l_{a=-1/2}$
($l=1,2,3,\cdots$).\cite{Kishimoto:2002xi} \ 
Using the operator $\hat A$, we have demonstrated that
the one-loop vacuum energy at the solution is independent of 
moduli such as interbrane distances.
These results are consistent with the interpretation that
the solution represents the tachyon vacuum.
We have also found a homotopy operator for another type of 
identity-based solution whose associated function has higher order
zeros.\cite{Igarashi:2005wh} \ 
We can apply the same procedure to prove $\delta Z(t)=0$
(\ref{eq:dZt=0}) for this solution and 
a particular type of solution in the $KBc$
subalgebra,\cite{Schnabl:2010tb}
 which is a real form of an identity-based solution.

We have also revisited the cohomology problem for the identity-based
solutions.\cite{Kishimoto:2002xi,Igarashi:2005wh} \ 
Using the obtained homotopy operator, one can conclude that
there is no cohomology at all the ghost number sectors.
The nontrivial cohomology part of (\ref{eq:cohomology2}) 
(or (\ref{eq:cohomology2_4}))
cannot be regarded as a BRST exact state within a single Fock space.
This is not the first appearance of such a state in SFT.
In the bosonic closed light-cone SFT, a classical solution associated
with the dilaton vacuum expectation value was
constructed and then it was impossible to realize it
within a single Fock space.\cite{Yoneya:1987gc} \ 
Then, it was remarked that the space of string
fields should be much larger than a single Fock space.
Also, in the study of target space duality, it was emphasized that
classical solutions in SFT must live outside the Hilbert space of the
original background.\cite{Kugo:1992md} \ 
More recently, a tachyon vacuum solution based on wedge-like
states\cite{Schnabl:2005gv}
includes a so-called phantom state. The phantom state is effectively
zero in a Fock space, but 
it is indispensable to derive the vacuum energy correctly.
Once again, for the cohomology of $Q_l$, we are forced to incorporate
the state outside a single Hilbert space.

In the discussion so far, the anticommutation relation 
$\{Q_l,\hat A\}=1$ 
is respected on any state.
However, the relation (\ref{eq:bzkUl=0}) might imply that
$
\varphi=(Q_l\hat A+\hat A Q_l)\varphi
$
is not equal to
$Q_l(\hat A\varphi) +\hat A (Q_l\varphi)
$,
which gives zero for (\ref{eq:varphidef}),
if one interprets Eq.~(\ref{eq:bzkUl=0}) as it stands.
Namely, the associativity of multiplication of the operators 
may be broken on the states of the form (\ref{eq:varphidef})
because multiple infinite summations of oscillators
are included in the expression
and we have interchanged the order of limits naively.
In order to avoid this ambiguity, some regularizations should be
introduced and/or the space of states should be restricted
appropriately.
Together with the issue stated in the previous paragraph,
a mathematically more rigorous treatment of the space of states
in SFT is desired in future developments.
Although one can find some investigations in this direction in the
context of the $KBc$ subalgebra in Ref.~\citen{Schnabl:2010tb},
for example, a wider class of string fields should be incorporated to
resolve the delicate problems mentioned above.

Finally, we comment on the result for the one-loop vacuum energy 
at the tachyon vacuum from the viewpoint of the BRST quartet
mechanism.\cite{Kugo:1979gm,Hata:1980yr} \ 
In the present case, the BRST charge is the operator $Q_l$ and any
states are classified into the irreducible representations of the
algebra of $Q_l$ and the FP ghost charge.
The existence of the homotopy operator for $Q_l$
shows that there are no BRST singlet states.
Therefore, we might be able to interpret that the vanishing result of
the trace (\ref{eq:dZ1}) is closely related to the norm cancellation
among quartet states.
In contrast, if the anticommutation relation $\{Q_l,\,\hat{A}\}=1$ is
broken on the nontrivial part of (\ref{eq:cohomology2}),
the state (\ref{eq:varphidef}) seems to propagate in the trace
as a BRST singlet. However, the state cannot form a singlet pair with
a nonzero inner product.
In fact, its dual state in the trace should have the ghost number $2l+2$
or $2l+1$ and such a state does not belong to the BRST singlet
representation as seen in (\ref{eq:cohomology2}).
Therefore, the result $\delta Z(t)=0$ in \S \ref{sec:1loop} seems to be
plausible also from the above speculation.

\section*{Acknowledgements}
We would like to thank Masako Asano, Yuji Igarashi, Katsumi Itoh and
Haruhiko Terao for helpful comments and useful discussions.
The work of I.~K. and T.~T. is supported by
JSPS Grant-in-Aid for Scientific Research (C) (\#21540269).

%

\end{document}